\documentstyle[12pt]{article}
\begin{document}
\def\be{ \begin{equation}}
\def\ee{ \end{equation}	}

\title{On Heteropolymer Shape Dynamics}

\author{
Pawel Pliszka and Enzo Marinari$^{(*)}$\\
Dept. of Physics\\
Syracuse University\\
Syracuse, NY 13244, USA\\[0.4em]
{\small pliszka@suhep.phy.syr.edu marinari@roma2.infn.it}\\[1.0em]
$(a)$: and Dipartimento di Fisica and Infn, \\
Universit\`a di Roma { \em Tor Vergata}\\
Viale della Ricerca Scientifica\\
00173 Roma, Italy
}

\maketitle

\begin{abstract}

	We investigate the time evolution of the heteropolymer model
introduced by Iori, Marinari and Parisi to describe some of the
features of protein folding mechanisms. We study how the (folded)
shape of the chain evolves in time. We find that for short times the
mean square distance (squared) between chain configurations evolves
according to a power law, $D \sim t ^\nu$.  	We discuss the
influence of the quenched disorder (represented by the randomness
of the coupling constants in the Lennard-Jones 	potential) on value of
the critical exponent.  We find that $\nu$ decreases 	from
$\frac{2}{3}$ to $\frac{1}{2}$ when the strength of the quenched
disorder increases.

\end{abstract}

\vfill
{\hfill {\bf SCCS 330, hep-lat/9207011} }
\vfill

\newpage

\vfill
\newpage

\section{Introduction}

Modeling the protein folding is one of the most relevant challenges open to
Statistical Mechanics. The idea that relevant issues of the folding process can
rely on the disordered nature of the amino-acid system has been put forward and
investigated by many groups (see for example refs. \cite{OLD} and references
therein). Shakhnovic and Gutin$^{\cite{SHGU}}$ in their seminal work have been
giving a first quantitative form to these ideas, implementing the mechanism of
Parisi's Replica Symmetry Breaking$^{\cite{BOOK}}$. Mezard and Parisi have
applied a systematic RSB treatment to random manifolds$^{\cite{MEPA}}$.

Iori, Marinari and Parisi in ref. \cite{IMP} (IMP) have studied a very
simple heteropolymer model, attempting to emphasize the difference
between the static structure of spin glasses and native proteins. The
model turned out to present potential important features, like the
dominance of one or a small number of ground states. In ref.
\cite{FUKU} the model is studied in $d=2$, and ref. \cite{MAP2}
studies the case of Lennard-Jones homopolymers.

The Hamiltonian of the IMP model (describing a self-interacting heteropolymer)
has the form

\be
  H= \sum _{i,j;i \ne j} \bigl [
  h d_{i,j}^2 \delta_{i,j+1}
	 - \frac{A}{d_{i,j}^6}
	 + \frac{R}{d_{i,j}^12}
	 + \frac{\sqrt{\epsilon} \eta_{i,j}}{d_{i,j}^6} \bigr ] \ ,
  \protect\label{EHAM}
\ee

where $i,j$ range from $1$ to $N$ (the number of {\em sites} of the
polymeric chain), and $d_{i,j}^2 \equiv (r_i-r_j)^2$. The first term
represents a harmonic force holding the adjacent sites together. This
term prevents the chain from breaking into pieces. The contributions
proportional to $A$ and $R$ (attractive and repulsive) form a
conventional Lennard-Jones (L-J) potential.  The last term is a
quenched disorder contribution; it models a nonuniformity of the
dipoles interacting by the L-J potential. The values of $\eta _{i,j}$
are independent and uncorrelated stochastic variables (which do not
vary in time).  The $\eta _{i,j}$'s have mean value zero and variance
one. The set of coefficients $\eta _{i,j}$ can be viewed as
characterizing a single native protein.

The most relevant questions concern typical realizations of the heteropolymeric
chains. In the following we will mainly interested in quantities which are
averaged both on the thermal noise and on different realizations of the
$\eta _{i,j}$ couplings.

In this note we investigate the short time behavior of the
heteropolymeric chains whose dynamics is defined by the Hamiltonian
(\ref{EHAM}).  Specifically, we study the diffusion of the distance
between two chain configurations defined as

\begin{equation}
  D_2(\alpha,\beta) \equiv {1\over N^2} \sum_{i,j}
	 ( d_{i,j}^{(\alpha)} - d_{i,j}^{(\beta) })^2\ ,
  \protect\label{EDIS}
\end{equation}

where $\alpha $ and $\beta $ label two chain configurations at
different instances of time.  The time evolution of $D_2$ tells us
about the diffusion of the shape (and size) of the folded chain.  We
concentrate our attention on the dependence of diffusion on
$\epsilon$, the quenched noise strength.

Analytical studies of the diffusion in the (simple) case of polymer
where the interaction is local along the chain are possible in the
context of normal mode decomposition of the Langevin dynamics (LD).
Such simple models are known to yield a characteristic power law
dependence of the mean-square displacement (squared) over
time$^{\cite{Doi}}$, $t^{\frac{1}{2}}$.  This power law holds in the
range of times where the finite size effects are still invisible, but
collective effects are already important$^{\cite{Doi}}$. This is an
interesting regime to which we will dedicate attention in the
following.  Other relevant models are discussed in the context of the
problem of surface aggregation (see for example \cite{Medina} and
\cite{Kardar}). For example for a nonlinear model with a local
interaction in $(1+1)$ dimensions the mean square displacement
(squared) grows as $t^{\frac{2}{3}}$ in the relevant
regime$^{\cite{Kardar,Plischke}}$.

In the case of interest (heteropolymers with long range interaction on
the chain, and a L-J potential with a random quenched attractive
contribution) the problem is very complex (since the Hamiltonian is
disordered and non-local, and frustration can play a role). It is also
important to remind that we want to study a situation where $N$ is
large but not infinite, and this finiteness can play a role. This is
true both for our model (we will present here results obtained with
$N=30$) and for the potential application to protein folding. In a
biological application of our model the sites of our chain have to be
identified with (pre-assembled) parts of the secondary structure, and
$N$ would be of order $10$. The heteropolymeric chain is finite and
localized (at equilibrium the distance between chain sites is finite:
translational and rotational degrees of freedom do not enter in our
definition of mean displacement). This implies that, after a time
large enough, the mean square displacement reaches a maximum (which,
roughly speaking, characterizes the amplitude of thermal fluctuations
and depends on spatial extent of the chain).

Nonlinear systems with external bias are studied numerically and are
expected to yield crossover between $t^{1/2}$ and $t^{2/3}$ power
laws.  It is not completely clear whether the apparent nonlocality of
interaction can be treated effectively as a bias (an effective mean
field due to cloud of particles) incorporated in a local nonlinear
model describing nearest-neighbor interaction along the chain.

\section{The Equilibrium Shape}

We have tried to get some more informations about the equilibrium
shape of the IMP heteropolymers. We have mainly used in our runs the
same parameters used in IMP$^{\cite{IMP}}$.

Let us look at some raw number in order to try to understand what happens when
we go from the homopolymer phase to the {\em folded} phases, where the disorder
plays a crucial role.  In absence of the quenched disorder ($\epsilon=0$), for
$\beta=1$, $N=30$, $a=3.8$, $r=2$ and $h=1$, the total chain energy is
$<\!E_T\!>$ $=$ $-273.0 \pm .5$,
the average chain first neighbor square length
$<\!d^2_{i,i+1}\!>$ $=$ $1.55 \pm .10$ ,
and the end to end chain length
$<\!d^2_{N,1}\!>$   $=$ $6.0 \pm 1.5$.
The corresponding values for $\epsilon=6$ are
$<\!E_T\!>=-635 \pm 50$,
$<\!d^2_{i,i+1}\!>$ $=$ $1.7 \pm 0.5$
$<\!d^2_{N,1}\!>$   $=$ $ 3.5 \pm 3.5$
(here the errors are computed by averaging over different realizations of the
$\eta_{i,j}$ couplings). These number are here only to hint an order
of  magnitude. Indeed in ref.~\cite{IMP} it has been seen that
$100$ millions of full chain sweeps are not at all sufficient
to explore the entire phase space.

The values of $\epsilon$ needed in IMP (for $N=30$) to go deep in the
folded phase ($\epsilon \approx 6$) are quite large.  We have then to
be quite careful in checking which terms are contributing to the
ground state total energy. We find that in both the disordered and in
the $\epsilon=0$ regimes the attractive part of the L-J energy
expectation value (in which we include the part proportional to
$\epsilon$) is roughly equal to twice the repulsive part of the
energy. This is what we find for the minimum of a L-J potential for a
two-particle system, where at equilibrium

\be
  \frac{A}{d^6} = \frac{2R}{d^{12}}
\ee

($d$ is the distance between the two particles at equilibrium).  For
these values of the parameters the harmonic term contributes about ten
percent to $<\!E_T\!>$ but practically does not affects the value of
$d$.

For $\epsilon=0$ in the limit $N \to \infty$ the ground state
configuration is such that the chain sites lie on a face-centered
cubic lattice. The lattice (which in this case coincides with the
solution to the {\em kissing problem}, i.e. finding the maximum number
of spheres of diameter $d$ tangent to a given one) can be seen as made
of centers of spheres of diameter $d$ packed regularly with the
maximum density (which in $3d$ is $12$).  In this limit the harmonic
energy term only determines the order in which the chain sites are
placed on the lattice sites. Fukugita, Lancaster and
Mitchard$^{\cite{FUKU}}$ have shown that in $d=2$ the lattice
structure is quite clear in the $\epsilon=0$ case, and survives
(although the evidence of ref. \cite{FUKU} is in this case less
compelling) the transition to the disordered phase. In the $3d$ case
we find that for $N=30$, in the {\em folded} phase, the lattice
structure is lost. We clearly see the lattice structure in the
$\epsilon=0$ case, and we see that it becomes weaker but partly
survives when thermalizing the chain in the strongly disordered
potential~\cite{IMPL}.

\section{The Time Dependence of the Shape}

In order to discuss the diffusion of chain configuration
quantitatively, it is more convenient to introduce the following
definition of distance between two configurations

\begin{equation}
  D_4(\alpha,\beta) \equiv {1\over N^2} \sum_{i,j}
	 ( {d^2} _{i,j}^{(\alpha)} - {d^2}_{i,j}^{(\beta) })^2\ ,
  \protect\label{EDIS4}
\end{equation}

Since

\be
 ( {d}_{i,j}^{(\alpha)2} - {d}_{i,j}^{(\beta)2 }) =
 ( d_{i,j}^{(\alpha)} - d_{i,j}^{(\beta) })
 ( d_{i,j}^{(\alpha)} + d_{i,j}^{(\beta) }) \ ,
\ee

where the second factor is much larger than the first, the exponent
characterizing the power law short-time dependence of $D_2$ and $D_4$
is the same.  One could also use the two
definitions of the distance introduced in ref. \cite{IMP}, and expect
to find the same critical behavior. We study the diffusion of the
polymer shape by analyzing an ensemble average of $D_4$ as a function
of the time separation between the two configurations. We average over
the time dynamics and over different realizations of the $\eta_{i,j}$
couplings, and compute

\be
  D_4(t) \equiv <\!D_4(\alpha(t),\alpha(0))\!> \ ,
\ee

where $ \alpha(0)$ is an equilibrium configuration.

In all our runs we have used a standard Monte Carlo dynamics.

The fluctuations of the critical exponent governing the behavior of $D_4(t)$
from sample to sample are quite small. We average $D_4$ over a few hundreds
($100$-$500$) of starting points for a given sample.

In order to check that we are really picking up the correct universal behavior
we have studied a simple harmonic chain. Let us stress that the problem is very
delicate: there is a non universal short time region (where the discreteness of
the Metropolis procedure plays an important role), an asymptotic constant value
for $D_2$ (with some universal approach to $D_4$, and in between the
(short) time region we are interested in. We have to check that we are
observing, in this region, a true scaling behavior, and we can be sure of that
only in the limit of large $N$. In order to check our results for the simple
harmonic chain we have compared them to the analytic expression resulting from
the normal mode expansion of the Langevin dynamics (LD).

The Langevin equation for the harmonic chain is of the form

\begin{equation}
  \zeta \frac{dr_i}{dt}=-k \ (2r_i-r_{i+1}-r_{i-1}) +f_i \ ,
\end{equation}

where the uncorrelated  random forces $f_j$ satisfy

\be
  <\!f_i(t_1) f_j(t_2)\!>=2\
  \zeta\  k_B\ T\  \delta(t_1-t_2)\  \delta_{i,j}\ .
\ee

$\zeta$, a friction constant, sets here the time scale for the problem.
In the continuum limit  $(2r_i-r_{i+1}-r_{i-1})$ may be replaced by
$\partial^2 r / \partial n^2$. One can introduce normal modes in the
standard way by

\be
  x_p= \frac{1}{N} \int_0 ^N  dn \ r(n) \ \cos(\frac{p \pi n } {N})
\ee

Using Wick's theorem for evaluating average of products of the
coordinates $<\!x_j x_k x_m x_n\!>$ one can easily compute the
$D_4(t)$, finding

\be
  D_4(t)=\sum_{i,j} \bigl [ \kappa_{i,j}(0) -\kappa_{i,j}(t) \bigr ] \  ,
\protect\label{D4SUM}
\ee

where

\be
  \kappa_{i,j} (t)= c \ \{ \sum _p  e^{-p^2 t / \tau_r}
	 { 1\over p^2}
	 { \bigl (\cos ({p \pi i \over N}) - \cos (
	 {p\pi j \over N}) \bigr ) ^2} \} ^ 2 \ ,
\ee

and

\be
  c=64 \times 4N\frac{b^2}{\pi^2}= 64\times {4N 3k_B T\over h};
\ee

\be
  \tau_r={N^2 b^2 \over kT 3 \pi^2} =
 	{\zeta N^2 \over h \pi^2}\ .
\ee

$b^2=kT/3h$ is the asymptotic equilibrium expectation value of $d_{i,i+1}^2$.
We show the comparison of our numerical fitted data with the analytic result
in fig. $1$. We can see very well that the power for short times is
$\frac{1}{2}$. Saturation starts to show up for larger times (and it
is very clear for very large times, not included in the figure), also
if the statistical error is already becoming, in this regime, very large.

Let us just remind again that we are using a discontinuous dynamics, the
Metropolis algorithm. We know that we are getting the correct asymptotic
equilibrium distribution, but for short times we do not have a simple
correspondence with the corresponding Langevin dynamics. Let us discuss this
point in some detail. Consider a Metropolis dynamics, where we propose
 a trial
random increment defined by a displacement vector $\vec{\delta}$
chosen from some probability distribution $P(\vec{\delta})$.
The increment is accepted with probability
$P_a= {\rm min}(1,e^{\beta (H(\vec{r})-H(\vec{r}+ \vec{\delta}))})$.
Therefore the  average of a single step displacement is
\be
  <\!\vec{r'} -\vec{r}\!> \quad \sim \quad <\! \vec{\delta} \
	\rm{min}(1,e^{\beta (H(\vec{r})-H(\vec{r}+ \vec{\delta}))})\!>\ .
\ee

For small $\beta \vec{\delta} \cdot \vec{\nabla} H(r)$, (corresponding
to a large acceptance factor) the   $ <\!\vec{r'} -\vec{r}\!>$
 be approximated  as
\be
   -\beta \>  <\!  \vec{\delta}
\> \Theta(\vec{\delta} \cdot \vec{\nabla} H(\vec{r})) \>
  (\vec{\delta} \cdot \vec{\nabla} H(\vec{r}))\!>
 \ =\ - {\beta \over 6} <\! \delta^2 \!> \vec{ \nabla} H(r)   \ ,
  \protect\label{ESMA} \ee
where $\Theta$ is the step function.
This is the same form one gets for a Langevin step, with a scale
factor which depends on the acceptance ratio of the Metropolis
procedure, $<\!\delta^2 \!> \beta$.  Normally one uses the Metropolis
algorithm with a trial displacement that is not small: this is indeed
the big advantage of the Monte Carlo method.  That means that in
general condition (\ref{ESMA}) does not hold.  We expect the
Metropolis dynamics to reproduce the continuous dynamics only for
times larger than the typical time of continuous dynamics necessary to
reach the mean displacement $<\!\delta\!>$.  This is the time region we
are interested in, and that we discriminate before analyzing the
dynamical critical exponent. In this region (of times large enough not
to feel the non-universal details of the dynamics, but small enough
not to describe the relaxation to the asymptotic mean displacement) we
will see that we can determine a critical exponent $\nu$

\be
  D_4(t) \simeq t^\nu\ .
\ee

In the context of Langevin Dynamics we can relate the exponent $\nu$
with the  asymptotic behavior  of   dynamics linearized around local minima.
We consider the normal modes eigenvalues of the linearized dynamics,
$\lambda_p$.
Let us assume that asymptotically (i.e. in the continuum limit
of large $N$ and small $p$)

\be
  \lambda_p = p^{\alpha}
\ee

and that the fluctuations of all modes are the same and not
correlated.  The width of the probability distribution of the $p$-th
mode behaves as $\frac{C}{\sqrt{\lambda_p}}$, where $C$ is a mode
independent constant. Then we obtain

\be
  D_4(t)=\sum_{i,j} \bigl [ \kappa_{i,j}(0) -\kappa_{i,j}(t) \bigr ] \  ,
\ee

where

\be
  \kappa_{i,j} (t)= c \ \{ \sum _p  e^{-p^\alpha t / \tau_r}
	 { 1\over p^\alpha}
	 { \bigl ( c_{p,i}  - c_{p,j} \bigr ) ^2} \} ^ 2 \ ,
\ee

where the $c_{p,i}$ are the scalar products of the $p$-th mode vector
$x_p$ with the position vector of the $i$-th site, $r_i$.  Changing
the order of the summation and converting the sum over $p$ into an
integral (in the continuum limit) we obtain

\be
  D_4(t) \sim  \int _0 ^ \infty dp \quad p^{-\alpha}
  (1-e^{{t \over \tau_r} p^\alpha})
  \sim t ^{ 1- {1\over \alpha}} \qquad\ ,
\ee

where the coefficients $c_{p,i}$ were replaced by their average values
and summed over~\cite{Doi}.
Therefore, under the above approximation,
the expected relation between the exponent $\alpha_1$ and the
exponent $\nu$ is

\be
  \nu= 1-{1\over \alpha}\ .
\ee

We have verified numerically that in the large $\epsilon$ region the
above relation is actually satisfied.  This is consistent with the
fact that $\nu$ is a true universal exponent, dependent only on the
energy spectrum and independent from $\beta$.

Our numerical result do clearly exhibit, in the large $\epsilon$  region, the
scaling relation

\be
  \lambda_p \sim {(\frac{p}{N})}^\alpha\ .
\ee

for $N$ as small as  $15-30$, justifying the claim we are close to the
continuum limit
(obviously we ignore the first six eigenvalues which are zero  due to the
rotational and translational symmetries).

\section{Numerical Estimates of $\nu$}

We present here our numerical determination of the dynamical critical exponent
$\nu$ we have defined before, from the {\em short} (but not so short) time
behavior of $D_4$.

The four parameter which determine the Hamiltonian, $A$, $R$, $h$ and
$\epsilon$ affect the behavior of $D_4(t)$. Let us note, at first,
that the time scale $\tau_r$ after which the power law does not hold
any more

\be
  D_4(t) \simeq  k t^{\nu}\ , \   t << \tau_r
\ee

is a decreasing function of the strength of interaction, and an
increasing function of the system size. Secondly, the coefficient $k$
decreases with the strength of coupling	and increases with the
temperature and the system size.  In fig. $1$ we have shown the
behavior of an harmonic chain, together with the results of the
theoretical analysis.

In fig. $2$ we show the time behavior of ordered systems for two different
values of the parameters. Here (in the globular phase) we find a good fit with
$\nu=\frac{2}{3}$, which coincides with the prediction of models with local
interaction. The fact that this exponent stays the same even if we increase the
local interaction be increasing $h$  ten times while keeping $R$ and $A$ fixed
suggests that in the coil phase the diffusion properties in the absence of the
quenched noise can be adequately described by a model of sites with a local
interaction placed  in an effective external field.

In the following we will mainly be interested in the dependence of the exponent
$\nu$ over $\epsilon$.

Let us note at first that the exponent $\nu$ does not seem to depend on the
particular realization of the quenched noise. We have not performed large scale
simulations, which would be needed in order to get quantitative precise results
in systems with such a dramatic critical slowing down (see ref. \cite{IMP}),
but we have got from our simulations quite a precise qualitative picture.
We show in fig. $3$ $D_4(t)$ for three different realizations of the noise,
with a very similar power behavior. The non-universal coefficient of the power
behavior does indeed depend on the given noise realization, but $\nu$ does not.
Of course, the energy of interaction itself does not determine
$\nu$ (as it does not determine the equilibrium shape).

In fig. $4$ we show $D_4(t)$ for different $\epsilon$ values. For increasing
$\epsilon$ we observe a reduction  of $\nu$: for $\epsilon=10$
$\nu = .52 \pm .05$

As noted in ref. \cite{IMP} the correlation time shows a dramatic
dependence on $\epsilon$: in fig. $5$ we show the distribution
probability $P_t(D_4)$ for four values of $\epsilon$. We have selected
the same time moment for the $4$ distributions. The probability
$P_t(D_4)$ is obtained by averaging over several hundred initial
configurations. While for $\epsilon =0$ the $P(D_4)$ for times $\simeq
10 \theta $ ( where $\theta \equiv 10^4$ updates of each chain site)
is nearly stationary, for $\epsilon=6$ times of two orders of
magnitude longer are may be just starting to be enough (see fig. $6$).
The new {\em Tempering} approach to Monte Carlo dynamics, recently
proposed in ref. \cite{TEMP}, has chances to alleviate the situation.

\section*{Acknowledgments}

We thank Sergei Esipov for his observations about the local
interaction and for an interesting discussion, Giulia Iori for
interesting discussions, for her comments concerning the lattice
structure and for providing us with the eigenvalue decomposition, and
Alan Middleton for very helpful interactions during the course of this
work.

\newpage

\section*{Figure Captions}

  \begin{description}

    \item {1}
      Comparison between the $D_4(t)$ (here $t_{MC}$ is the number of
full MC sweeps of the chain) estimated in our Monte Carlo runs for the
purely harmonic chain (scattered dots) and the theoretical prediction
from the Langevin dynamics normal mode expansion (Eq. \ref{D4SUM},
continuous curve). Here $\tau_r = 800$.

    \item {2}
      As in fig. $1$, but for a homopolymer ($\epsilon=0$) with
      Lennard-Jones interaction, with $A=3.8$ and $R=2$, log-log plot.
	The two curves are  for $h=1$ and for $h=10$.

    \item{3}
      $D_4(t)$ for three different realizations of the
      quenched noise $\eta_{i,j}$, $\epsilon=6$.\\
	a) linear plot, \ b) log-log plot (arbitrary normalization).

    \item{4}
      $\log(D_2)$ versus $\log(t)$
      for different values of $\epsilon$ (arbitrary normalization).
	The steepest line (labelled with {\em random})
        corresponds to a system of
	free particles - all proposed moves accepted.

    \item {5}
      The distribution probability $P_t(D_4)$ for
      four values of $\epsilon$. We take the same time
      for the $4$ distributions (twice the maximum time
      shown in fig. $4$).

    \item {6}
      The plot of time dependent probability distributions for
      $\epsilon=6$ for three different values of time. The smallest time $t$
      corresponds to the time of fig. $5$.

\end{description}
\end{document}